\documentclass[aps,prl,longbibliography,10pt,superscriptaddress,showpacs,showkeys,twocolumn,floatfix]{revtex4-1}
\usepackage{amssymb}
\usepackage{amsmath}
\usepackage{amsfonts}
\usepackage{graphicx}
\usepackage{epstopdf}%\usepackage{color}
\providecommand{\revision}[1]{#1}
\newcommand{\deleted}[1]{}

\begin{document}
\newcommand{\De}{{\mathrm D}}
\newcommand{\hw}{\hbar\,\omega}
\newcommand{\nbr}{n_\b}
\newcommand{\Ea}{E_\mathrm{a}} %\newcommand{\K}{K^+}
\newcommand{\ii}{\mathrm{i}}
\newcommand{\Sb}{\text{Sb}}
\newcommand{\Ge}{\text{Ge}}
\newcommand{\cm}{\text{cm}}
\newcommand{\seg}{\text{s}}
\newcommand{\eV}{\text{eV}}
\newcommand{\nm}{\text{nm}}
\newcommand{\DB}{{\text{DB}}}
\newcommand{\ILM}{{\text{ILM}}}
\newcommand{\ph}{\text{ph}}
\renewcommand{\d}{\text{d}}
\newcommand{\kb}{k_B}
\newcommand{\RT}{\text{ph}}
\newcommand{\ann}{\text{ann}}
\newcommand{\app}{\text{app}}
\newcommand{\C}{$^\circ$C}
\newcommand{\Cmath}{^\circ\text{C}}
\newcommand{\juan}[1]{{#1}} %uncomment this command and comment the previous one to eliminate the Juan and bf.
%\begin{frontmatter}

\title{Infinite charge mobility in muscovite at 300K}%

\author{F. Michael Russell}
% \affiliation{Group of Nonlinear Physics, Universidad de Sevilla, ETSII, Avda Reina Mercedes s/n, 41012-Sevilla, Spain }
\email{mikerussell@us.es}

\author{Juan F. R. Archilla}
\affiliation{Group of Nonlinear Physics, Universidad de Sevilla, ETSII, Avda Reina Mercedes s/n, 41012-Sevilla, Spain}
\email{archilla@us.es}
\thanks{Corresponding author}
\author{Fabi\'an Frutos}
\affiliation{Group of Surfaces, Interfaces and Thin Films, Department of Applied Physics I, Universidad de
Sevilla -- ETSII, Avda Reina Mercedes s/n, 41012-Sevilla, Spain}

\author{Santiago Medina-Carrasco}
\affiliation{X-Ray Laboratory (CITIUS), Universidad de Sevilla -- Avda Reina Mercedes 4B,
41012-Sevilla, Spain}

\date{December 30, 2017}

 \begin{abstract}
 Evidence is presented for infinite charge mobility in natural crystals of muscovite mica at room temperature. Muscovite has a basic layered structure containing a flat monatomic sheet of potassium sandwiched between mirror silicate layers. It is an excellent electrical insulator. Studies of defects in muscovite crystals indicated that positive charge could propagate over great distances along atomic chains in the potassium sheets in absence of an applied electric potential. The charge moved in association with anharmonic lattice excitations that moved at about sonic speed and created by nuclear recoil of the radioactive isotope $^{40}$K. This was verified by measuring currents passing through crystals when irradiated with energetic alpha particles at room temperature. The charge propagated more than 1000 times the range of the alpha particles of average energy and 250 times the range of channelling particles of maximum energy. The range is limited only by size of the crystal.
\end{abstract}
%\begin{keyword}
%\sep DLTS \PACS

%\end{keyword}
%\end{frontmatter}
\maketitle

\section{Introduction}

\begin{figure*}[t]
\begin{center}
\includegraphics[width=0.6\textwidth]{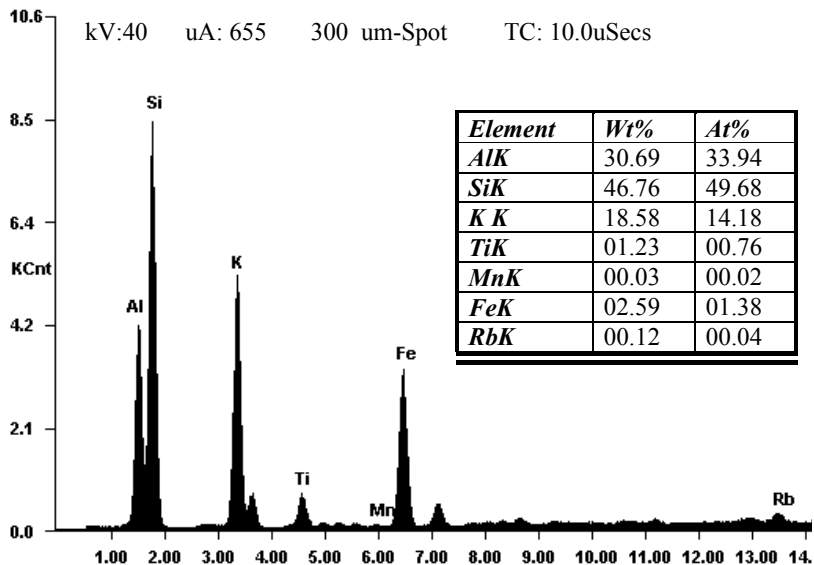}
\caption{X-ray fluorescence (XRF) spectrum of the sample in powder form and resulting composition of the sample used. It is consistent with typical natural crystals of muscovite. Note that elements with Z$<$11 are not detected with XRF, therefore O, F and H are absent from the table.}
\label{fluorescence} \end{center}
\end{figure*}
\begin{figure}[h]
\begin{center}
\includegraphics[width=\columnwidth]{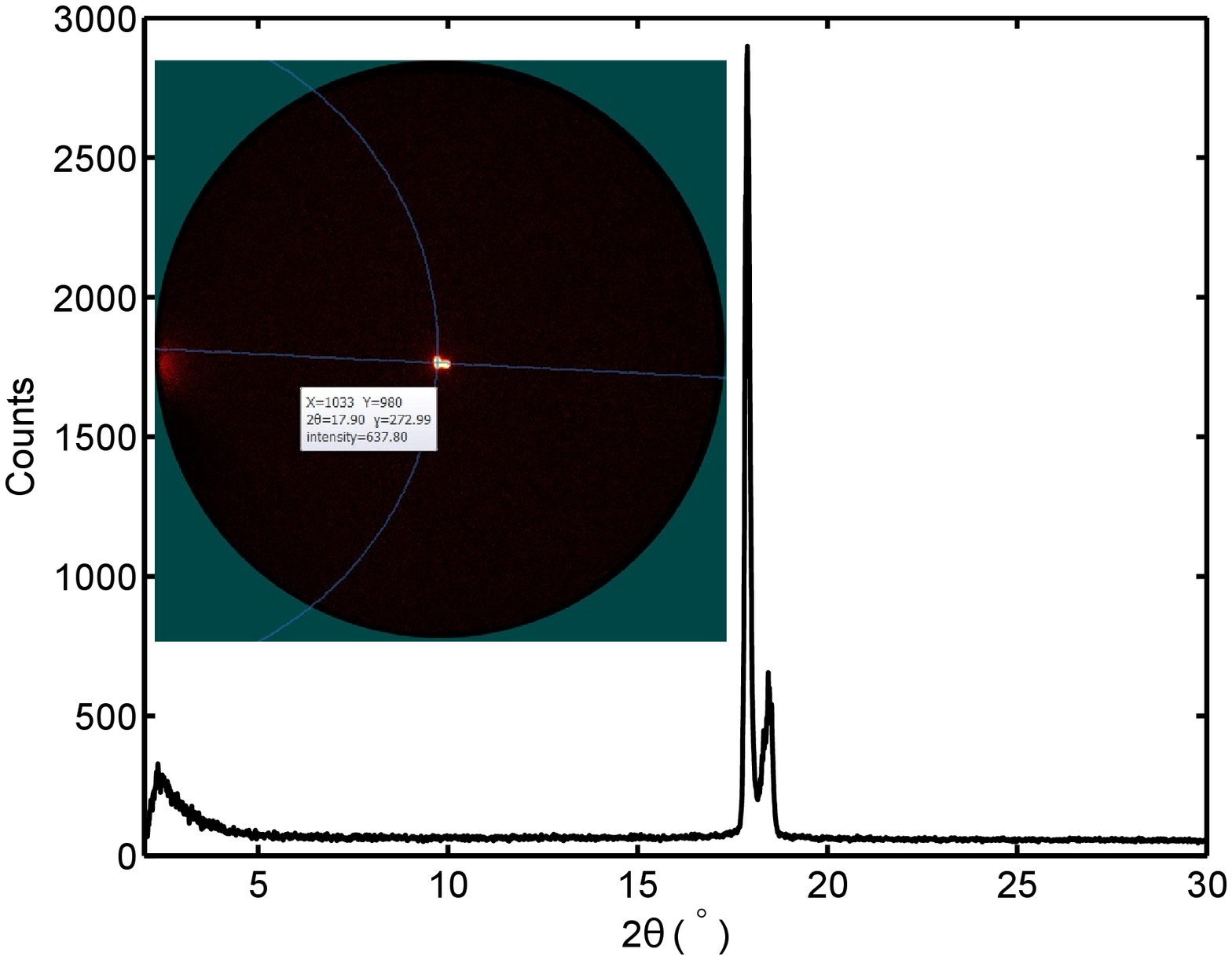}
\caption{(Color on line) X-ray diffractogram obtained on the flat side of the sample with an X-ray area detector Bruker V{\AA}NTEC-500 with $\lambda= 1.5406$\,\AA. The isolated maximum corresponding to an angle $2\theta=17.9^\circ$ indicates that the planes $\{001\}$ are parallel to the flat side with $d=h\lambda/(2\sin(\theta))\simeq20\AA$ with $h=4$. The frame of the area detector in the inset shows the isolated point characteristic of a layered single crystal. %The frame shows an angular $2\theta$ range of approximately $30^\circ$.
Two artifacts appear in the diffractogram: a signal at low angle and another at a slightly higher angle than the main peak. These are caused by the difficulty of assembling the sample exactly in the plane corresponding to the direction parallel to the (001) plane.}
\label{vantecwithframe}
\end{center}
\end{figure}

Muscovite mica is a layered mineral of composition K$_2$[Si$_6$Al$_6$O$_{20}$](OH,F)$_4$ with the potassium lying in monatomic sheets between mirror silicate layers. Clear sheets split from crystals of muscovite have high electrical resistivity of order $10^{13}$\,$\Omega$m at ambient temperature, persisting up to temperatures of order 1000\,K, above which crystals start to decompose. In the past it has been widely used as an insulator. Recently, new uses in electronics have been suggested through bandgap engineering\,\cite{kim-nanosheets2015}. Natural crystals containing significant amounts of impurity of iron become metastable as they cool after growth, when swift charged particles can trigger local precipitation of magnetite\,\cite{russell-tracks2015article,russell-positive1988}. The initial precipitation delineating the tracks continues until the metastable state ends. This accretion leads to permanently recorded fossil tracks that are visible~\cite{russell-tracks2015article,russell-identification1988}. These fossil tracks can have lengths of several centimetres. Some of them, due to their observed properties, can be attributed to high energy positrons emitted from the radioactive $^{40}$K isotope in muscovite\,\cite{russell-Optik1993}. Other fossil tracks lie in principal crystal directions and are associated with mobile anharmonic lattice excitations, called quodons, created by nuclei recoiling from decay of $^{40}$K\cite{russell-Optik1993}. The recording processes yielded magnetite or epidote depending on the crystal composition. These allowed the first and exclusive study of individual moving charges, anharmonic lattice excitations and their interactions in a crystal. Some of the quodon tracks exceed 30\,cm in length. The complexity of the unit cell of muscovite has hindered the understanding of the structure of quodons created by nuclear recoil. As a result it was not known if their structure was the origin of the positive charge that triggered the recording process. Measurements of the extent of decoration on quodon tracks showed that the extent of decoration was equal to that of slowly moving positrons\cite{russell-arxiv2015}. This led to the suggestion that quodons might trap and transport charge through the lattice in the potassium (001)-planes~\cite{russell-tracks2015article,archillaLoM2016}. Related studies of trapping of charge by mobile anharmonic excitations in 2D arrays added credence to this suggestion\cite{velarde2010,chetverikov2013,chetverikov2016}. Numerical modelling and analogue studies of simplified muscovite lattices showed that quodons could be longitudinal breathers~\cite{Marin1998,Russell-Collins1996,russell2011,bajars2015}. The propagation of breathers in other 2D lattices has also been described~\cite{korznikoza-hexagonal2017,wattis-honeycomb2014,hizhnyakov-graphene2016}.  The existence of quodons in muscovite was shown experimentally by producing them with alpha particles and detecting the ejection of atoms in the closest-packed directions\,\cite{russell-experiment2007}. The designation of quodons as breathers is compatible with the observed creation of multiple daughter quodons by scattering at lattice defects. Most types of anharmonic excitation move at supersonic speed\,\cite{archilla-kosevich-pre-2015} but breathers can be slightly subsonic, moving at about 3500\,m/s. As quodons can move in six possible directions in the muscovite (001)-plane there is no known source of electric potential that might assist the transport of charge in crystals. The absence of applied electric potentials across the crystals implied that the mobility of the charge, defined as   $\mu= v/E$ where $v$ is the speed of motion of the charge and $E$ is the applied electric field, when coupled to the breather is infinite. Here we present evidence for this effect from experiments conducted at room temperature on crystals of muscovite.

\begin{figure}
\begin{center}
\includegraphics[width=\columnwidth]{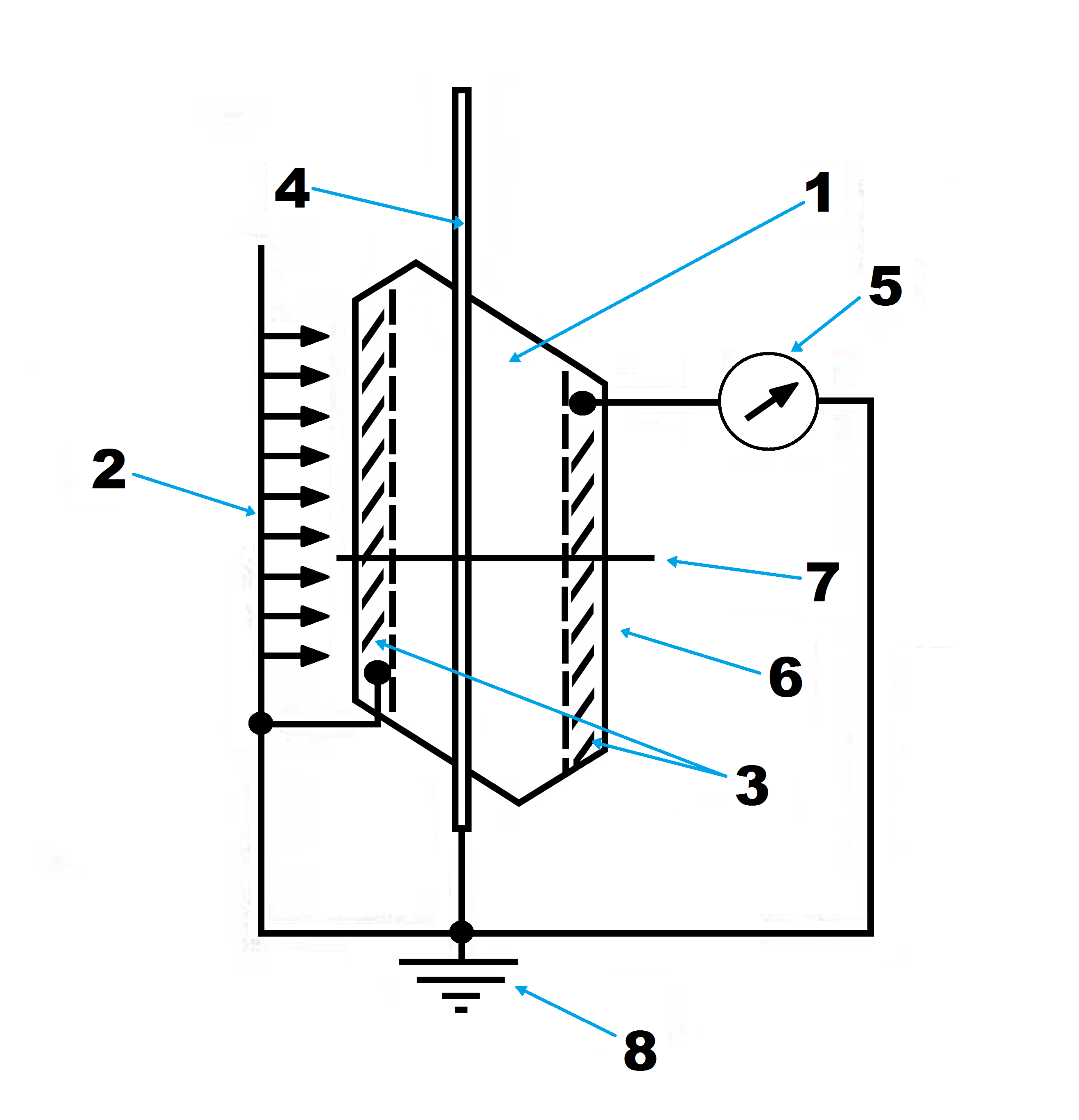}
\end{center}
\caption{Diagram of plan view of experiment. 1 is the crystal, 2 the alpha source, 3 the gold contacts, 4 the screen preventing air-borne currents, 5 the current meter, 6 the (010)-face of the crystal and 7 the [100]-crystal direction. \revision{Label 8 marks the electrical earth}. Dimensions are 2\,cm long, 1\,cm wide \revision{in the plane of the paper and 0.5\,mm thick in the perpendicular direction.}}
\label{setup-plan}
\end{figure}

\begin{figure}
\begin{center}
\includegraphics[width=\columnwidth]{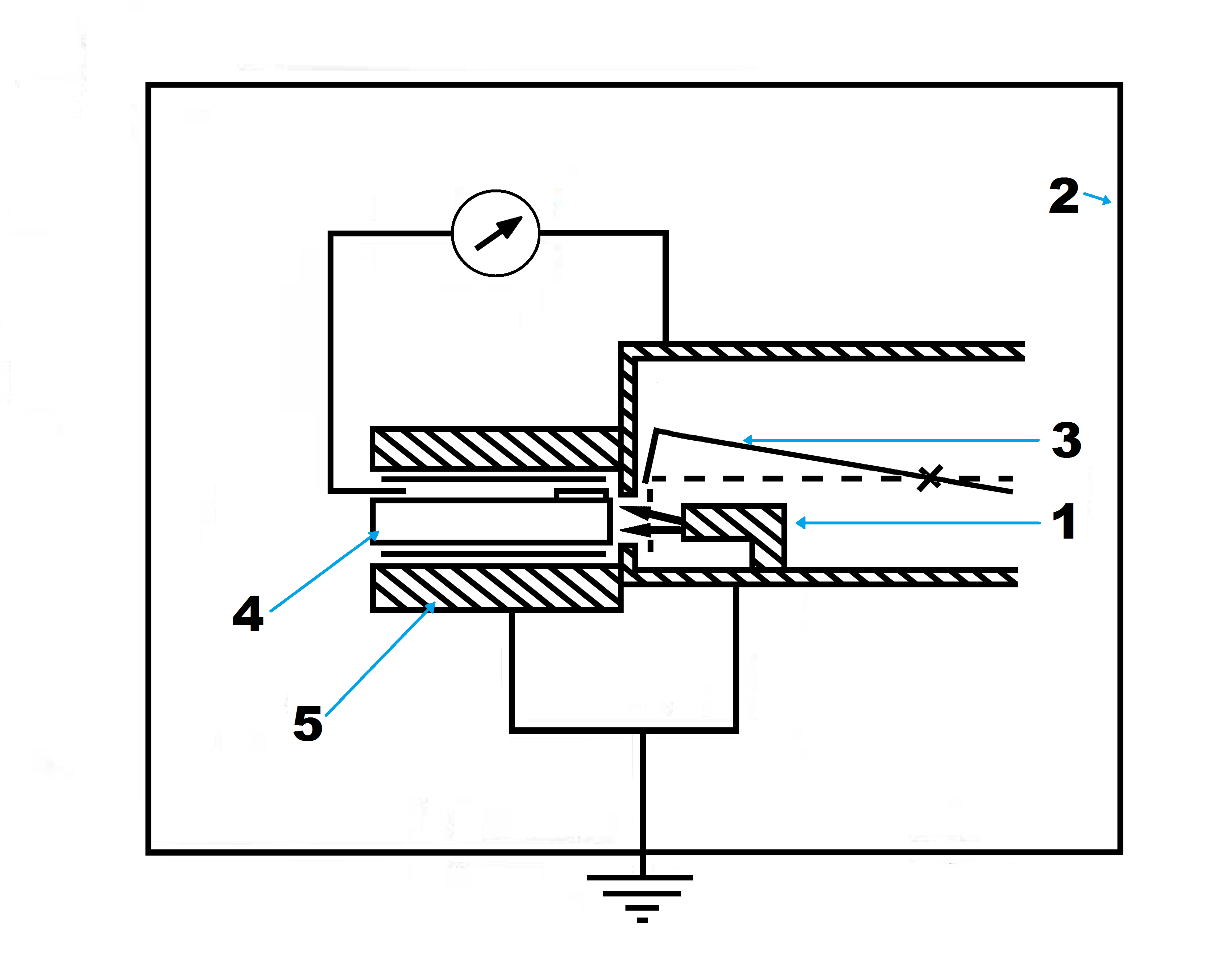}
\end{center}
\caption{Side view of the experiment. 1 is the mount for the alpha source, 2 the Faraday cage, 3 the movable shutter to interrupt the alpha beam and 4 the crystal. \revision{The crystal is held between two earthed metal plates 5 that are part of the Faraday cage. The crystal is insulated from the plates 5 by two PTFE sheets.}}
\label{setup-side}
\end{figure}

\section{Experimental setup}
\revision{The design of the experiment follows closely that used in the earlier experiment that demonstrated the existence of quodons\,\cite{russell-experiment2007}. That experiment also showed that the charges moved at the same speed as the quodons. Muscovite is an excellent electrical insulator so charge cannot move over macro distances in an undisturbed crystal. If the charge trapped by a quodon moved at a different speed to that of the quodon then it would escape from the quodon and become immobile, which is incompatible with the observation of long tracks with the same extent of decoration as slowly moving positrons or holes}

To study this hypothesis of charge propagation in muscovite by quodons required their creation and the existence of free charges. This was achieved by irradiating one edge of a well-formed crystal with 5\,MeV alpha particles from $^{241}$Am. The composition of the crystal was found by X-ray fluorescence (XRF)
  %juanpostprint XRD changed to XRF above
spectrometry of the sample in powder form; the result shown in Fig.~\ref{fluorescence} is consistent with a natural crystal of muscovite. The orientation of the crystal was determined using an X-ray area detector Bruker V{\AA}NTEC-500\,\cite{XRD2} as seen in Fig.~\ref{vantecwithframe}. The crystal was free of any precipitated magnetite or epidote, indicating that the recording processes had not been operative. On penetrating the crystal edge the alphas cause atomic collisions simulating nuclear recoils and produce free charges by localised ionisation. Each alpha also introduces two positive charges to the crystal. Two opposite edges of a crystal were vacuum coated with a thin gold film to provide electrical contacts. The geometry of the experiment, dimensions of the crystal, location of the electrical contacts and directions of principal crystallographic directions are shown in Fig.~\ref{setup-plan}. To minimise surface currents the crystal was bonded \revision{to insulating sheets and held in the metal frame 5.} This frame was fixed to an earthed metal plate that provided a barrier against air-borne ionisation currents. The alpha source was enclosed in a metal box with a narrow gated-window that could be operated remotely. The arrangement of the gated source is shown in Fig.~\ref{setup-side}. The alphas hit the 2\,cm long edge of the crystal. Although the nominal activity of the source was 5\,$\mu$Ci, measurements at the source opening with a Geiger counter, combined with  the geometry of the experimental setup allowed to estimate the number of alpha particles impinging on the crystal as $\Phi_\alpha=3.5\times 10^3$\,alphas/sec per the 0.5\,mm thickness and 2\,cm long edge of the crystal. Some alphas were lost at the gold film. Figure~\ref{setup-side} defines the direction terms.  The crystal was placed directly in front of the window with the gold film connected to the metal box, which was earthed. A Keithley 617 electrometer was used to measure the current and sign of charge transported through the crystal.

\begin{figure}[t]
\begin{center}
\includegraphics[width=\columnwidth]{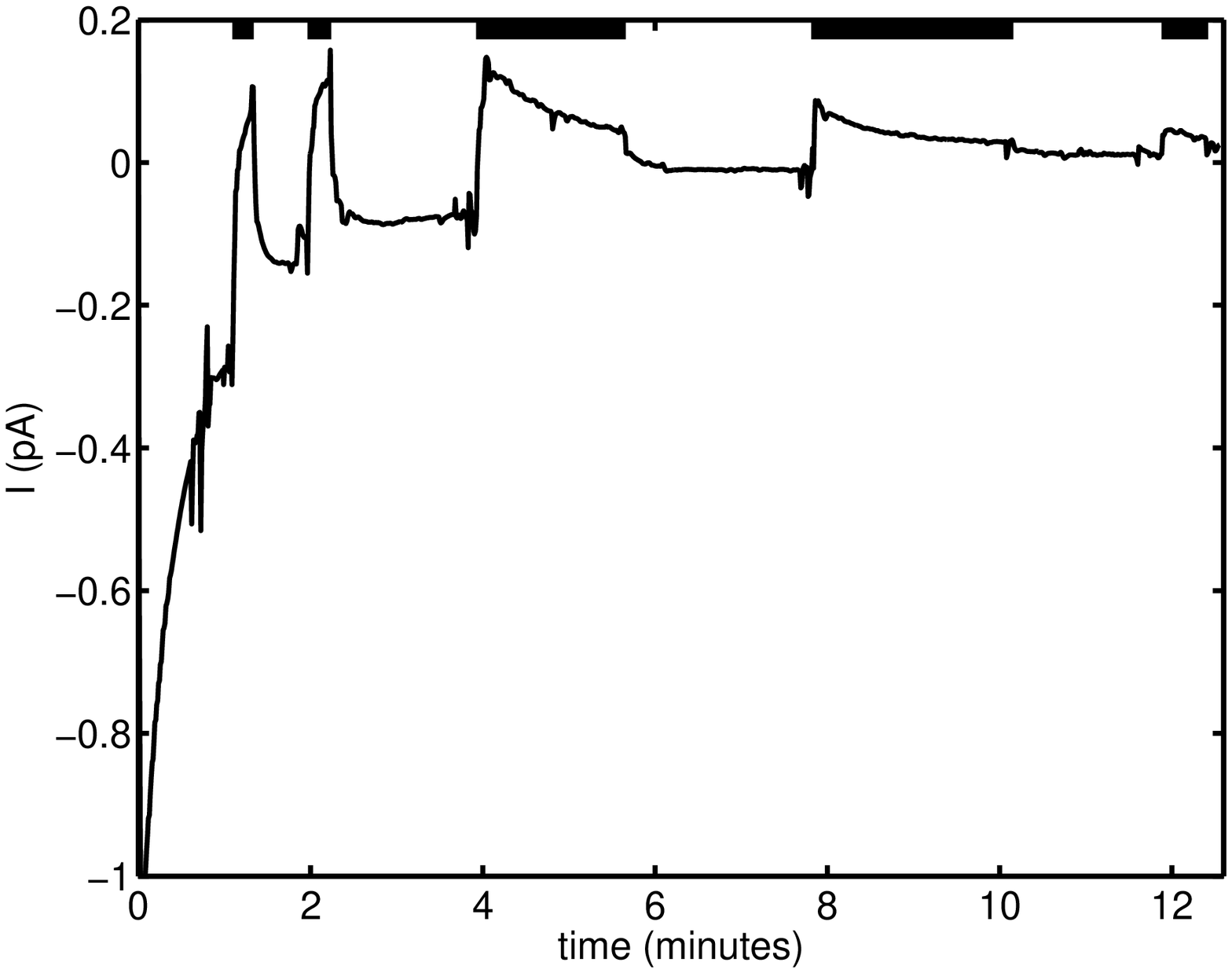}\\
\includegraphics[width=\columnwidth]{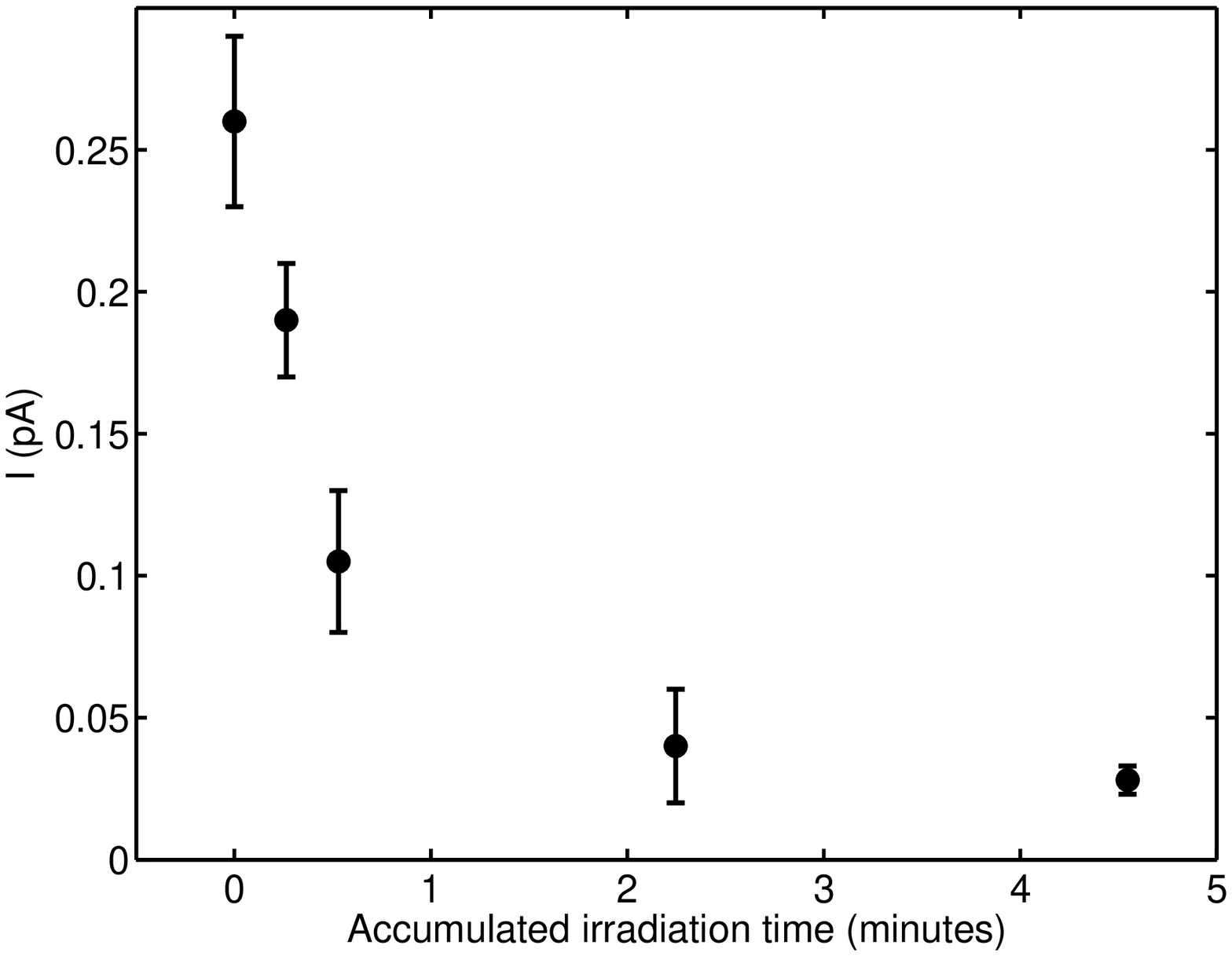}
\end{center}
\caption{\revision{{\bf Upper panel:}} Plot of the current measured by the Keithley pico-amp meter. The alpha irradiation was interrupted by closing the gate. It shows the large initial current expected due to charge build-up in the crystal and its decay as the creation of quodons continued. \revision{Black stripes at the top indicate alpha irradiation. {\bf Lower panel:} Plot of the measured average currents for the alpha irradiation periods with respect to the accumulated alpha irradiation time.}}
\label{fig_keithley}
\end{figure}

\begin{figure}
\begin{center}
\includegraphics[width=0.9\columnwidth]{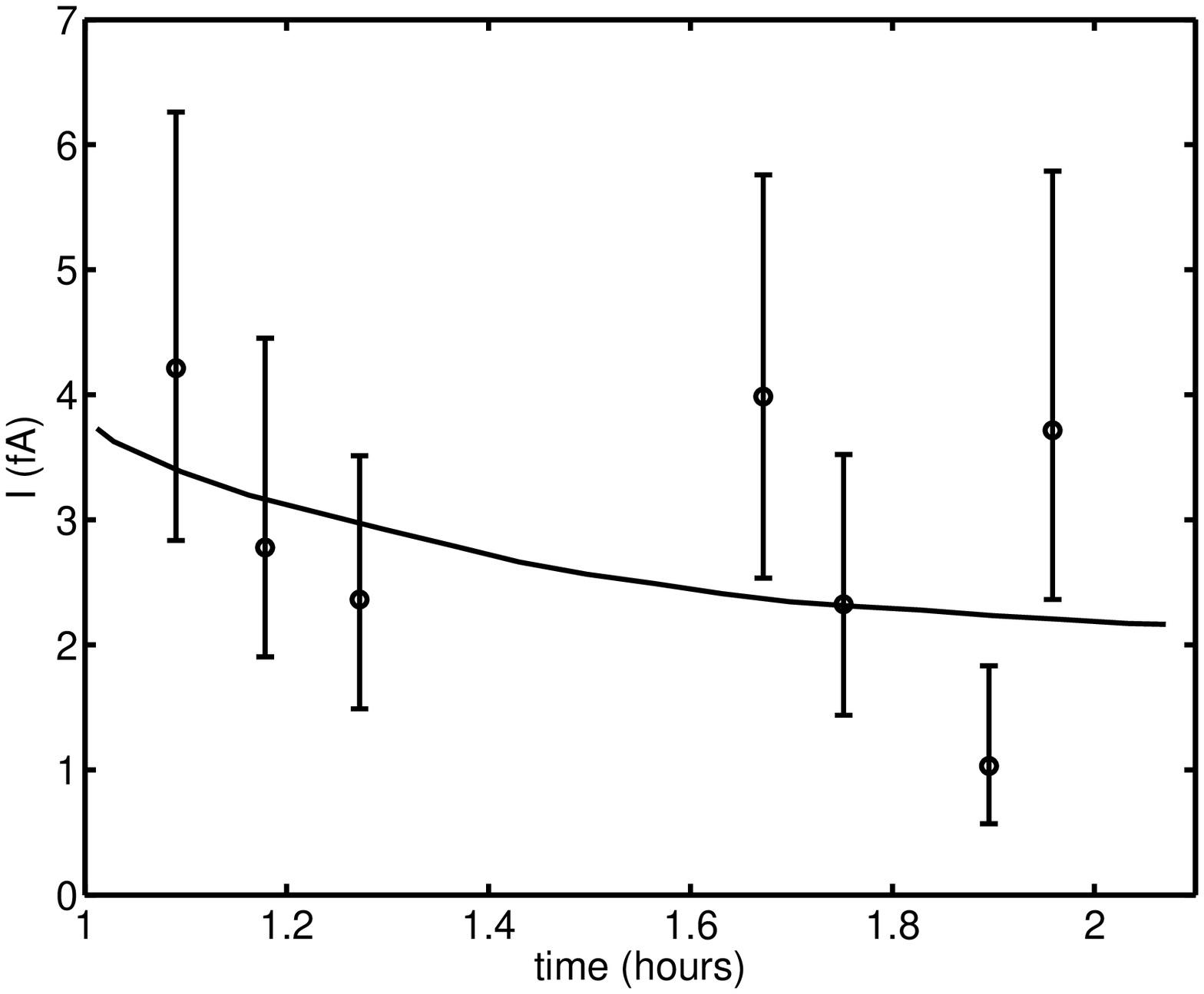}
\end{center}
\caption{Plot of the approach to the limiting current set by the rate of charge injected by the alpha source. The alphas source was too weak to give good signal to noise data. \revision{In absence of the alpha beam the current was $1 \pm 0.5$\,fA.} }
\label{Ilimit-sevilla-femto}
\end{figure}

\section{Results}
%The
%electrometer output for a typical exposure of the crystal to the alpha particles is shown in Fig.~\ref{fig_keithley}. The irradiation was pulsed to demonstrate the %dependence of the current on the alpha flux.

The electrometer output for the exposure of the crystal to the alpha particles is shown in Fig.~\ref{fig_keithley}. \revision{The currents were expected to be tiny and so random spikes due to various causes were anticipated, such as small vibrations of contacts at cables, possible movement of ionised or damp air or even cosmic rays. After switching on the electrometer a diminishing negative current flows which, according to the instrument manual\,\cite{keithleymanual}, stems from a combination of ground loop, electrostatic interference, thermal EMF, radio frequency interference, leakage resistance and input capacitance effects. Hence, the typical initial curve is a quasi-exponential negative current with random small spikes. However, against this background the effect of the alpha source being open or closed is easily discernible, as shown in Fig.~\ref{fig_keithley} (upper panel). The approximate magnitudes of the currents can be measured and are shown in the lower panel of the same figure with respect to the accumulated time of alpha irradiation. It shows that the currents are progressively smaller with increasing irradiation dose. This is consistent with the expected depletion of a reservoir of charge in the sample. The accumulated transfer of charge is about\,24 pC. The initial currents observed in the experiment could not be repeated because the charge reservoir became depleted and is rebuilt only very slowly as explained below.}

\begin{figure}
\begin{center}
\includegraphics[width=0.9\columnwidth]{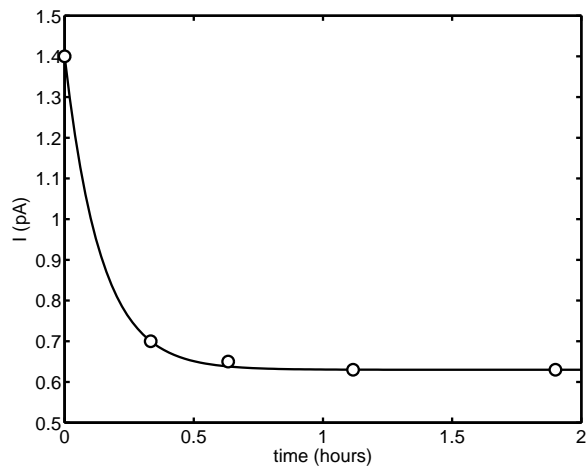}
\end{center}
\caption{Plot of the approach to limiting current with a stronger alpha source for a thicker section of the same crystal. \revision{The current with no alpha beam was $\simeq 0.02\pm 0.005$\,pA.}}
\label{Ilimit-uk-picoA}
\end{figure}

Filters of aluminium and muscovite of different thickness were inserted between the source and the crystal to investigate background effects. No current flowed when a foil of aluminium of 30$\mu$ thickness or 20$\mu$ of muscovite  was inserted, showing that the current was not due to X-rays. To examine background effects a test was performed with clear soda-lime-glass replacing the muscovite crystal. Although small transient impulses of current were observed when the source gate was operated no current that correlated with alpha irradiation was observed. \deleted{The upward curve of the electrometer output in absence of irradiation was due to instrument warm-up.} The measured current above background for extended irradiation times shown in Fig.~\ref{Ilimit-sevilla-femto} suggests an asymptotic limiting value under continuous irradiation of about 2\,fA, \revision{which is about 1\,fA above the current measured without alpha irradiation.} The existence of a limiting current was confirmed in a second experiment by using a stronger source with better source-to-crystal geometry and a thicker crystal. Figure~\ref{Ilimit-uk-picoA} shows the approach to a limiting current of 630\,fA in this second test.

\section{Analysis}
The dominant decay channel with 89.25\% probability of electron emission from $^{40}$K  leaves the daughter Ca$^{++}$ ion in the K layer\,\cite{radionuclides2012,archillaspringer2015article}. The electron scatters into the surrounding lattice. As muscovite is a good insulator this leads to a reservoir of holes and electrons. Diffraction scattering of the positrons concentrates them in the K-planes where holes remain following positron annihilation.  Pairs of free charges are created by the decay of potassium at the rate of about $5\times10^5$/cm$^3$\cite{radionuclides2012} per day and is expected to reach a saturation volume density limited by background radiation providing recombination paths and by internal conduction due to impurities.
\revision{From the volume of the sample, the decay rate and the observed charge depletion in the experiment, it can be obtained a time of about four years for the charge reservoir to recover its initial value. Quodons created by irradiation will pass through this reservoir and could capture them in flight.

When the radiation is turned on, it brings about a peak in the current. The first peak is the largest because the charge reservoir is full. Subsequent openings of the alpha gate led to smaller currents because the charge reservoir becomes progressively depleted. Also the current tends to decrease within each peak.}

\deleted{When quodons are randomly created they might approach and capture a free charge in flight. This should result in an initial peak in observed current.}

With further creation of quodons the reservoir of free charges continues to decrease and the current should asymptotically approach the rate of charge input by the alpha particles. Owing to the small activity of the $^{241}$Am source used in the first test the contribution of two positive charges per alpha to the crystal was expected to give a limiting current of $I_\textrm{limit}=\Phi_\alpha\times 2e\simeq 1$\,fA for the 0.5\,mm thickness of the crystal. \revision{As explained above, the observed limiting current was $\simeq 1$\,fA which is in good agreement with the current transported by the alpha particles within the many uncertainties of the experiment.} Following a period of regeneration of the charge reservoir under non-irradiation conditions a partial return to the initial peak current is expected and was observed.

The initial observed peak current was $I_\textrm{peak}=0.27$\,pA. For it to flow through the crystal a sufficient number of quodons had to be created. The charge need not flow only along chains in the [100]-direction but could also move along chains at  $\pi$/3 either side in a percolation manner,\cite{bajars2015}. If a quodon can trap only one unit of charge then the required number is at least $\dot{Q}=I_\textrm{peak}/e=1.6\times10^6$/sec. Since $\Phi_\alpha=3.5\times10^3$ alphas/sec impinged on the 0.5\,mm thick  and 2\,cm long the crystal  each alpha needed to create at least $\dot{Q}/\Phi_\alpha=450$ quodons. Assuming alphas had average energy of half the maximum of 5.4\,MeV then  the maximum number created of 20\,eV energy would be about $10^5$, which is more than two orders of magnitude greater than necessary. The extent of decoration on quodon tracks created by $^{40}$K decay as they lose energy by creating daughter quodons remains constant, showing that trapping persists over a wide range of energies, possibly as low as 1\,eV. Unlike phonons quodons have momentum, which must be supplied by the alpha particles. Alpha particles of 5.4\,MeV have sufficient momentum to create several hundred 20\,eV quodons rising to thousands of lower energy.
As quodons have momentum their interaction with the lattice via their de Broglie wavelength might influence their propagation and evident tolerance of lattice defects.

When working with small currents short transient spikes are inevitable when any change is made to the system, some due to contact potentials. Another concern is the possible build-up of voltage across a crystal due to injected charge but this was not possible in the experiment as current left the crystal.
%In principle, the voltage initially would be zero and then increase in proportion to the injected current. In the experiment the voltage across the crystal might %increase at the rate of 0.5\,V per minute if no current left the crystal.
In the experiment the earthed gold film on the irradiated edge prevented any voltage build-up. Internal ionisation of the crystal is overall charge neutral. Lastly, the large initial currents observed negate the possibility that they are due to electronic conduction.

The number of \revision{atoms} per unit volume in muscovite is about $10^3$ larger than in air, therefore the range of alphas in muscovite can be estimated as $10^3$ times smaller than in air which is 1-4\,cm for energies 2-5.5\,MeV. This means that the range of quodons is about 1000 times longer than the penetration of alphas of average energies and 250 times longer than the alphas of maximum energy.

\section{Conclusion}
\deleted{Studies of defects in natural crystals of muscovite identified a unique process that precipitated impurities in response to motion of positively charged particles and mobile, highly localised, anharmonic lattice excitations. This enabled studies to be made of positrons ejected from the decay of $^{40}$K in muscovite and of mobile lattice excitations, called quodons, created by associated nuclear recoils. Measurements on the fossil tracks showed that the charge on the quodons was equal to that of positrons. This indicated that charge could be transported through a crystal over great distances with minimal loss in absence of an applied electric potential. This prediction was verified by experiments at room temperature. This process of charge transport involving single charges is termed hyper-conduction (HC) to distinguish it from superconductivity. It is expected that this effect would persist up to much higher temperatures.}

\revision{Following studies of the extent of decoration on tracks of positrons and quodons in muscovite crystals it was predicted that charge could be transported through crystals of muscovite by quodons in absence of an applied electric potential field. This prediction has now been verified in the experiment described in this article. Alpha particles impinged on a sample of muscovite and produced a current about 200 hundred times larger than that injected by alpha particles, which was also measured. The initial high current observed was consistent with quodons capturing charge from a reservoir of charge slowly built up within the crystal due to $^{40}K$ decay. Although experimental confirmation of the existence of quodons, as predicted by the fossil tracks, had already been obtained, only recently was it deduced that they transport charge as shown by this experiment at room temperature. It provides further confirmation of the reality of quodons and assists in exploring their properties. Since the fossil tracks observed in muscovite crystals of quodons carrying a charge were recorded when the temperature of the crystals exceeded 700K it is possible that the transport of charge in suitable cables might also occur at elevated temperatures.} This process of charge transport involving single charges is termed hyper-conduction (HC) to distinguish it from superconductivity. \revision{We are presently studying the potential technological applications of this phenomenon.}

%% here a revision

%\revision{Insert here the text.
%See fig.~\ref{fig.1}, table~\ref{tab.1} and eq.~(\ref{eq.1}).
%See also~\cite{b.a,b.b}.}
%here a shortcut $\emc$ and again $\emc$
%\begin{equation}
%\label{eq.1}
%0\neq1
%\end{equation}

\acknowledgments
We wish to acknowledge useful discussions and help in interpreting these experiments by J. C. Eilbeck, L. Cruzeiro and L.  Brizhik.  FMR wishes to thank R. W. Witty for support and encouragement. JFRA wishes to thank project FIS2015-65998-C2-2-P from MINECO, Spain \revision{and a travel grant from VI PPIT-US from the Universidad of Sevilla, Spain}. All authors acknowledge IMUS for help in travel and organization.
%\nocite{*}
%\bibliography{bibrussellepl}
%\end{document}
\end{document}